\documentclass[journal]{IEEEtran}

\usepackage[noadjust]{cite}
\usepackage{filecontents}
\usepackage{gensymb}
\usepackage{graphicx}
\usepackage{epstopdf}
\usepackage{color}
\usepackage{placeins}
\usepackage{float}
\usepackage{tabularx,colortbl}
\usepackage{cite}
\usepackage{float}
\usepackage[caption=false]{subfig}
\usepackage{dblfloatfix}
\usepackage{balance}
\usepackage[ruled,vlined]{algorithm2e}
\usepackage{textcomp}
\usepackage{amsmath,amssymb}

\begin{document}
	\bstctlcite{IEEEexample:BSTcontrol} 
	\title{Array Layout Optimization in a 24-Element 38-GHz Active Incoherent Millimeter-Wave Imaging System}
	\author{Jorge R. Colon-Berrios,~\IEEEmembership{Graduate Student Member,~IEEE}, Derek Luzano,~\IEEEmembership{Graduate Student Member,~IEEE}, Daniel Chen, and Jeffrey A. Nanzer,~\IEEEmembership{Senior Member,~IEEE}}%
	
\maketitle

	\begin{abstract}
		
		Active incoherent millimeter-wave (AIM) imaging is a recently developed technique that has been shown to generate fast millimeter-wave imaging using sparse apertures and Fourier domain sampling.		
		In these systems, spatial frequency sampling is determined by cross-correlation between antenna pairs, making array geometry an important aspect that dictates the field of view (FOV) and image quality. This work investigates the impact of array redundancy and spatial sampling diversity on AIM image reconstruction performance. We present a comparative study of three receive array configurations, including one simple circular design and two arrays obtained through optimization strategies designed to maximize unique spatial samples while preserving system resolution and FOV. Performance is evaluated using the image-domain metrics of structural similarity index (SSIM) and peak sidelobe level (PSL), enabling a quantitative assessment of reconstruction fidelity and artifact suppression. We perform experimental validation using a 38-GHz AIM imaging system, implementing a 24-element receive array within a 48-position reconfigurable aperture. Results demonstrate that optimized array configurations improve spatial sampling efficiency and yield measurable gains in reconstruction quality compared to a conventional circular array, highlighting the importance of array design for AIM imaging systems.

	\end{abstract}
	

	\IEEEpeerreviewmaketitle
	
	\section{Introduction}
		
	Millimeter-wave imaging has emerged as a promising sensing modality for applications ranging from security screening~\cite{942570} and remote monitoring~\cite{yujiri2003passive} to autonomous perception and smart environments~\cite{4337827}, driven by its ability to provide non-contact scene characterization under diverse operating conditions. As imaging systems evolve toward wider deployment and commercial prevalence, fundamental challenges arise in achieving high-quality reconstruction while maintaining constraints on hardware complexity, calibration burden, and system scalability. Across array-based imaging architectures, these challenges are fundamentally linked to the efficiency with which spatial information describing a scene is sampled, as incomplete or redundant spatial frequency measurements can directly degrade reconstruction fidelity and limit achievable resolution. Although dynamic measurement strategies can provide additional scene information~\cite{11236959}, array geometry and measurement diversity~\cite{ulander1998ultra,rau2011multi,diebold2021unified,11271319} remain central design considerations in millimeter-wave imaging systems. This has motivated the development of approaches to maximize spatial sampling efficiency while avoiding increases in hardware complexity or resource requirements.

	Active incoherent millimeter-wave (AIM) imaging is an approach for reducing hardware complexity while preserving the ability to reconstruct spatial information through array-based sensing. By employing noise transmissions and interferometric processing, AIM systems form images through the cross-correlation of signals received across antenna pairs, enabling scene reconstruction without the need for fully coherent channel architectures~\cite{8458190,8904834}. In this framework, each antenna pair contributes a measurement of the scene spatial frequency content, making the distribution of antenna baselines an important parameter for reconstruction quality. As a result, the geometry of the receive array directly determines the diversity and quantity of spatial frequency samples, and thus the array design is a critical factor impacting image resolution, field of view, and artifact behavior in AIM imaging systems.
	
	Although interferometric imaging benefits from the availability of multiple antenna pairs~\cite{Thompson2001}, the resulting set of measurements often exhibits significant redundancy due to repeated baseline separations inherent to structured array geometries. Such redundancy leads to duplicated spatial frequency samples that, while potentially improving measurement robustness, does not increase the diversity of spatial information available for reconstruction. Imaging performance becomes constrained not only by the number of antenna elements but also by the distribution of unique baselines spanning the effective aperture. Insufficient spatial sampling diversity can manifest in the image as elevated sidelobe levels and image artifacts, and degrading the image quality. These observations highlight a fundamental design tradeoff between measurement redundancy and spatial frequency coverage~\cite{8861700,9624943}, motivating array design methodologies that explicitly consider spatial sampling efficiency as a primary optimization objective.
	
	Interferometric imaging is based on the van Cittert–Zernike theorem, which establishes that the cross-correlation of spatio-temporally incoherent signals received by antenna pairs provide samples of the scene visibility function in the spatial frequency domain. The distribution of antenna baselines defines the specific sampling of this Fourier space and directly impacts reconstruction performance. Antenna array optimization has been widely investigated across electromagnetic sensing applications~\cite{4512143,5525009,9769540}, with prior efforts focusing on objectives such as sidelobe reduction~\cite{768786,0749d8e86a124a648470521f844f7ec1}, aperture gain improvement~\cite{7305316}, bandwidth enhancement, and multi-objective design strategies~\cite{10416353}~\cite{996017}. These studies show the importance of array design optimization for improving the quality of images. This presents an opportunity for array design methodologies that treat spatial sampling efficiency in AIM imaging systems where sample diversity strongly influences reconstruction quality.

	In this work, we investigate the impact of spatial sampling diversity on AIM imaging through the design and evaluation of optimized receive array configurations. We consider three 24-element array geometries based on a defined grid of 48 possible element locations: one simple circular layout, and two layouts obtained via optimization strategies aimed at maximizing the number of unique spatial frequency samples while preserving system resolution and field of view. The resulting arrays are analyzed in terms of baseline redundancy, spatial frequency coverage, and image reconstruction performance. Quantitative evaluation is performed using structural similarity index (SSIM) and peak sidelobe level (PSL) as metrics to characterize reconstruction fidelity and artifact behavior. Experimental validation using a 38-GHz AIM imaging platform implementing a 24-element reconfigurable receive array within a 48-position aperture is performed. The presented study demonstrates that constrained optimization of the array layout improves spatial sampling efficiency and leads to measurable gains in AIM imaging performance.

\section{AIM Imaging}	

AIM imaging relies on interferometric processing to reconstruct the received signals. In this approach, information is captured in the spatial frequency (Fourier) domain using an antenna array, and signals from each pair of antennas are cross-correlated. According to the Van Cittert–Zernike theorem, widely used in radio astronomy, if the signals arriving at a pair of antennas are spatio-temporally incoherent, their cross-correlation provides a sample of the scene visibility function, given by
\begin{equation}
	\textcolor{black}{ \mathcal{V} \left(u,v\right) = \left<E_i (t)E_j^*(t)\right> }
\end{equation}
\noindent
 $\left<E_i(t)E_j^*(t)\right>$ represents the correlation of the electric fields received at antennas $i$ and $j$, located at $(x_i,y_i)$ and $(x_j,y_j)$, respectively. Each antenna pair contributes visibility samples at coordinates $(u,v)~=~((x_i~-~x_j)/\lambda,~(y_i-y_j)/\lambda)$ in the Fourier domain, where $\lambda$ is the wavelength. Due to the two-dimensional (2D) nature of the array, each pair provides multiple measurements: one along the baseline, its conjugate along the reverse baseline, and the DC component corresponding to autocorrelation of elements. These measurements together populate the visibility plane and determine the spatial frequency information available for reconstruction. The goal is to reconstruct the intensity distribution $I(\alpha,\beta)$ of the scene, where $\alpha=\sin\theta\cos\phi$ and $\beta=\sin\theta\sin\phi$ are the 2D direction cosines. The Fourier transform of this intensity yields the visibility function, which contains all the spatial frequencies necessary for perfect reconstruction. However, the number of spatial frequency samples that can be collected is limited by the number of antenna pairs (or baselines) in the array. Cross-correlating each pair forms a sampling function given by
\begin{equation} \label{Suv}
	S(u,v) = \sum_m^M \sum_n^N{\delta(u - u_n)\delta(v - v_m)}
\end{equation}
\noindent
where $N \times M$ represents the total number of possible samples. Antenna pairs that are farther apart capture higher spatial frequencies, while closer pairs capture lower spatial frequencies. Because the sampling function is defined on a $\lambda$-scaled grid, two antenna pairs with the same separation and orientation produce the same visibility sample, which we refer to as a redundant sample.

The sampled visibility acquired by the imaging system is given by
\begin{equation}
	\mathcal{V}_s = S \times \mathcal{V}
\end{equation}
\noindent
The reconstructed scene intensity can then be obtained via an inverse Fourier transform
\begin{equation}
	I_r(\alpha,\beta) = \sum_{n=1}^N \sum_{m=1}^M \mathcal{V}_s(u_n,v_m) e^{-j 2 \pi (u_n \alpha + v_m \beta)}
\end{equation}
\noindent
The reconstructed intensity can be interpreted in the spatial domain as the point spread function (PSF) of the imager convolved with the true intensity. The PSF is obtained with an inverse Fourier transform of the sampling function
\begin{equation}\label{psf}
	\mathrm{PSF} = \mathcal{F}^{-1}\{S\}
\end{equation}
The PSF provides insight into the quality of the imaging system. In an ideal scenario, the PSF would be a perfect point (an impulse), corresponding to complete Fourier coverage and perfect reconstruction. In practice, because only a finite number of antenna pairs are available, the PSF is broader and contains sidelobes. 

Millimeter-wave Fourier domain imaging is a process implemented on the receiving array, and via the Van Cittert-Zernike theorem the signals generated by the scene must be spatially and temporally incoherent. Passive imaging systems thus rely on capturing intrinsic thermal radiation from the scene, which are inherently incoherent across space and time for natural objects. Thermal radiation in the millimeter-wave region of the electromagnetic spectrum is exceedingly low in power, and thus passive imagers generally require significant receiver gains, often on the order of 100~dB or more. In contrast, AIM imaging imparts spatial and temporal incoherence on the scene by illuminating the scene with noise transmitters~\cite{8458190}. The transmitting array consists of independent noise sources arranged at wider spacings than the receivers with sufficient bandwidth to ensure spatial and temporal incoherence (see, e.g.,~\cite{9127123}). Actively transmitting signals generates signals scattered from the scene that are significantly higher in power than intrinsic thermal radiation. Therefore, receiving arrays can be implemented with typical receiver gains on the order of 20~dB, significantly reducing cost and receiver complexity.

\begin{figure}[t!]
	\noindent
	\centering
	\includegraphics[width=0.40\textwidth]{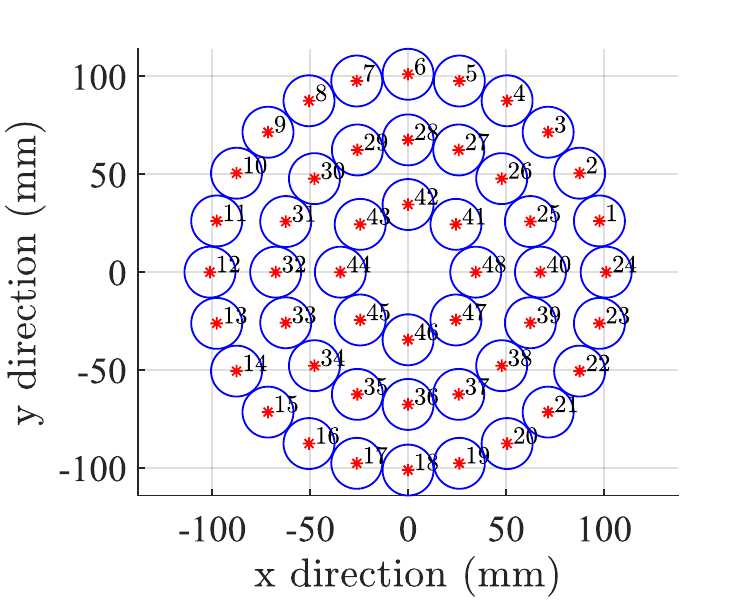}
	
	\caption{Possible spaces numbered for the multi-objective optimization. This is a representation of the physical possible spaces in the bracket that was designed for the imaging system. Circles with diameter equal to waveguide size are added around each possible placement, indicating the minimum inter-element distance constraint between elements constraint in the receive array. }
	\label{pos}
\end{figure}


\begin{figure*}[t] 
	\centering
	\includegraphics[width=0.3\textwidth]{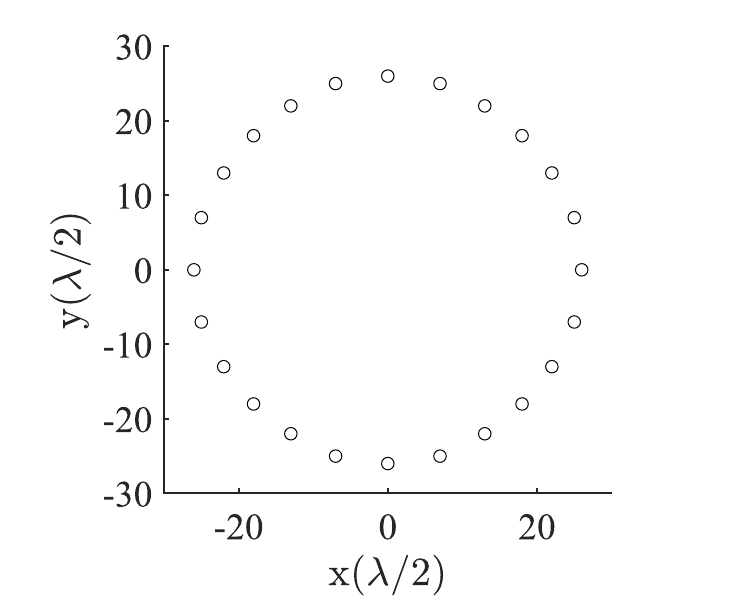}
	\hfill
	\includegraphics[width=0.3\textwidth]{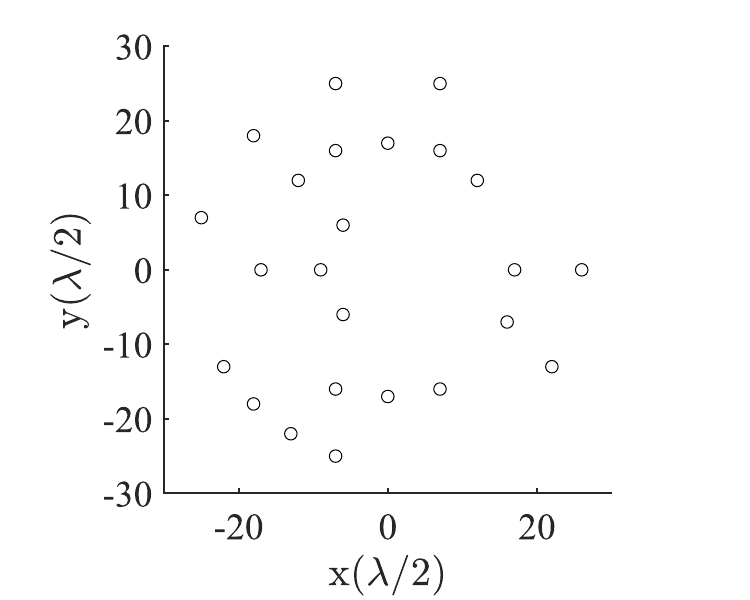}
	\hfill
	\includegraphics[width=0.3\textwidth]{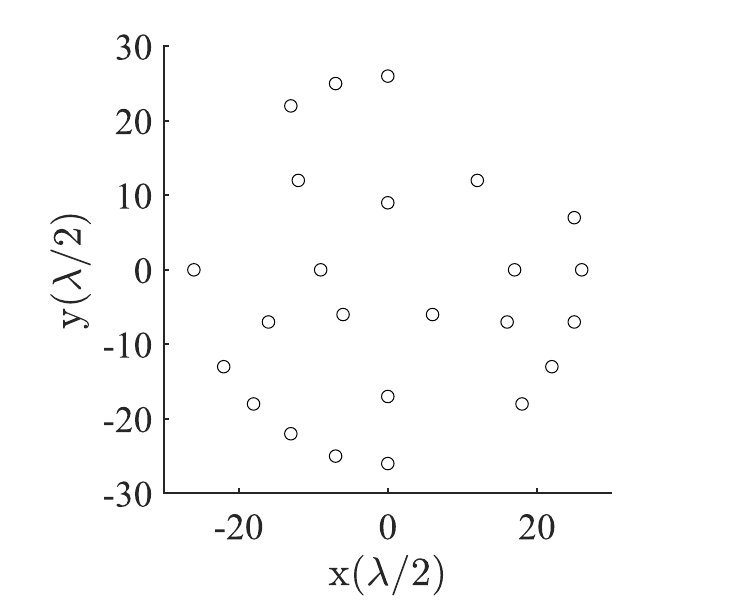}
	\caption{Array formations in $\lambda/2$ grid, (a) circular array, (b) random-search array, (c) MO Array.}
	\label{SimAr}
\end{figure*}
\begin{figure*}[t] 
	\centering
	\includegraphics[width=0.3\textwidth]{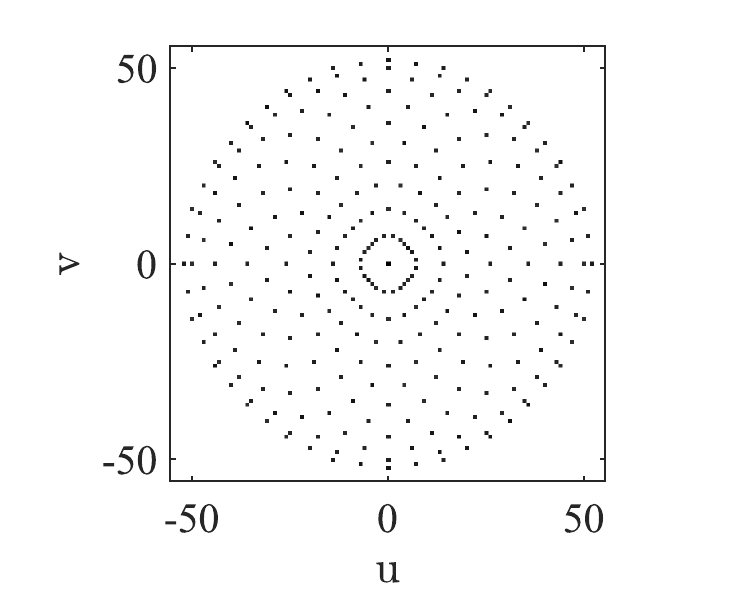}
	\hfill
	\includegraphics[width=0.3\textwidth]{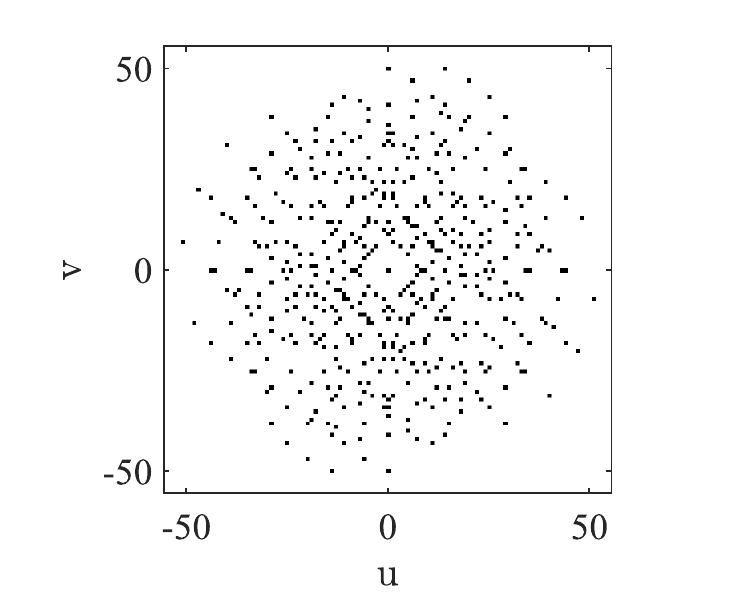}
	\hfill
	\includegraphics[width=0.3\textwidth]{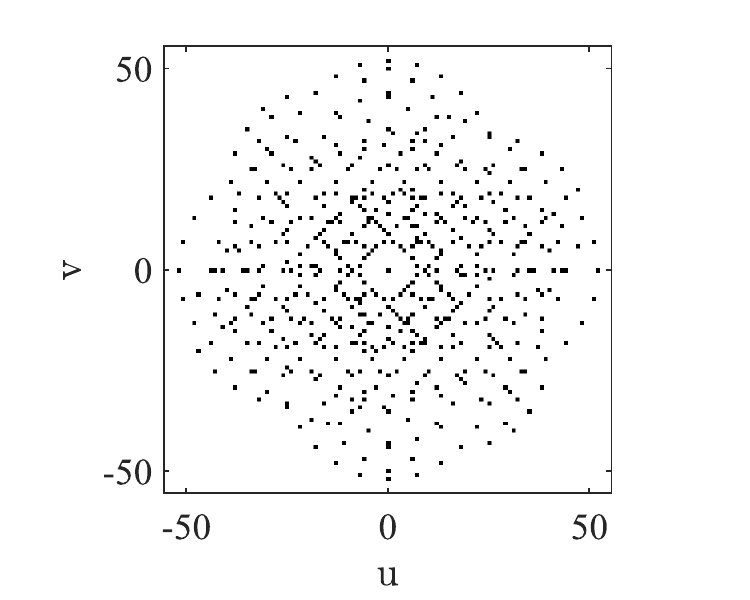}
	\caption{Sampling function of each array shown in figure~\ref{SimAr}.}
	\label{samp}
\end{figure*}

\section{Array Layout Optimization Approach}

The reconstructed image is highly dependent on the number of spatial frequency samples obtained by the array, and thus the array layout is a critical design aspect. Ideally, each antenna pair would contribute a unique spatial frequency sample; however in practice it is challenging to obtain array layouts with minimal baseline redundancy, particularly in two-dimensional layouts~\cite{4738328}. For any array, reduced redundancy can be achieved by implementing optimization algorithms. In this work we explore the optimization of an array layout using a constrained two-dimensional area commensurate with the array space available, minimum inter-element spacing of 26~mm and maximum of 202~mm. Prior AIM demonstrations used various numbers of elements, including 16-element arrays~\cite{9079644} and 24-element arrays~\cite{9753682}. To facilitate the study in this work, an array optimization framework with 48 possible position constraints was developed. The framework enables the optimization of AIM array geometries while constraining minimum and maximum baseline limits, which ensures physically realizable array layouts while improving the diversity of spatial frequency sampling. While this work focuses on a specific array configuration, the proposed approach is generalized and can easily be extended to other array formations.

The array consists of 24 possible analog to digital converters (ADC) channels connected to 15-dBi 3D-printed horn antennas that can be positioned across 48 possible locations, as illustrated in Fig.~\ref{pos}. Based on \eqref{Suv}, the sampling function of the system is obtained through cross-correlation of antenna pairs. Antenna pairs with identical baseline distance and orientation produce redundant samples. The sampling space can be optimized by strategically placing the 24 antennas within the 48 available positions to maximize sample coverage while reducing redundancy. A 2D array with 24 elements can produce up to $24^2 = 576$ unique samples. The antenna array design is optimized according to three criteria. First, sample redundancy is minimized to improve coverage of the visibility function in the spatial frequency domain, which corresponds to maximizing the number of unique samples in $S$. Second, the system resolution is optimized to enable clear image reconstruction with fine spatial detail. For an interferometric imager, the spatial resolution in the azimuth and elevation directions can be expressed in terms of the half-power beamwidth $\theta_{\mathrm{HPBW}}$ of the sinc-squared response associated with the largest baselines $D_x$ and $D_y$ along the horizontal and vertical axes of the array, respectively, as given by~\cite{Nanzer2012}
	\begin{equation}\label{res2}
		\Delta \theta_{\alpha,\beta} \approx \theta^{(\alpha,\beta)}_{\mathrm{HPBW}} \approx 0.88 \frac{\lambda}{D_{x,y}} .
	\end{equation}
Minimizing (\ref{res2}) leads to improved spatial resolution. Finally, the unambiguous field of view (FOV) is maximized to allow imaging of a larger scene region. The FOV is determined by the smallest inter-element spacings $d_x$ and $d_y$ along the horizontal and vertical axes and can be expressed in terms of the direction cosines $\alpha$ and $\beta$ as
	\begin{equation}\label{fov}
		\mathrm{FOV}_{\frac{\alpha}{2},\frac{\beta}{2}} = \frac{\lambda}{2 d_{x,y}} \, .
	\end{equation}
	
	\begin{figure*}[t!] 
	\centering
	\includegraphics[width=0.3\textwidth]{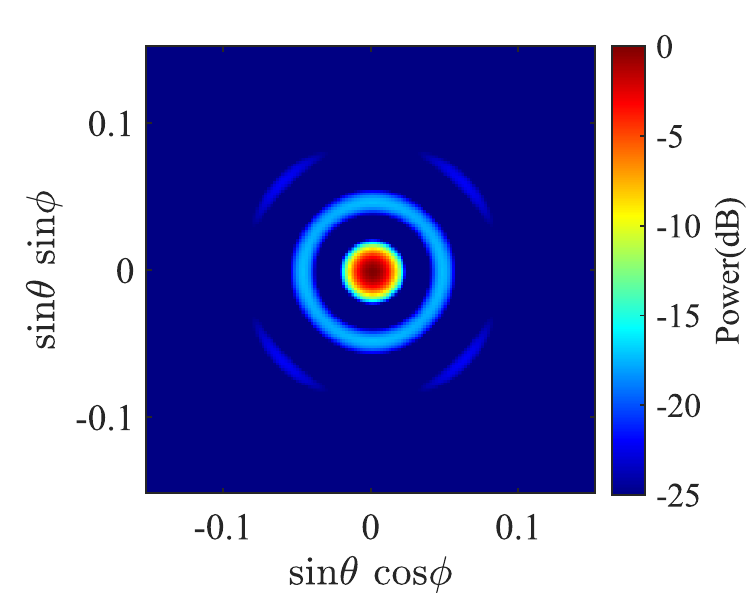}
	\hfill
	\includegraphics[width=0.3\textwidth]{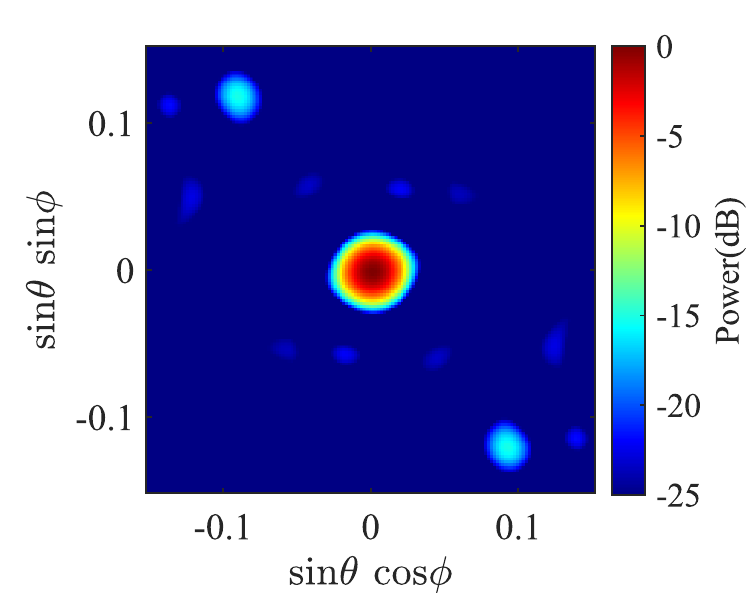}
	\hfill
	\includegraphics[width=0.3\textwidth]{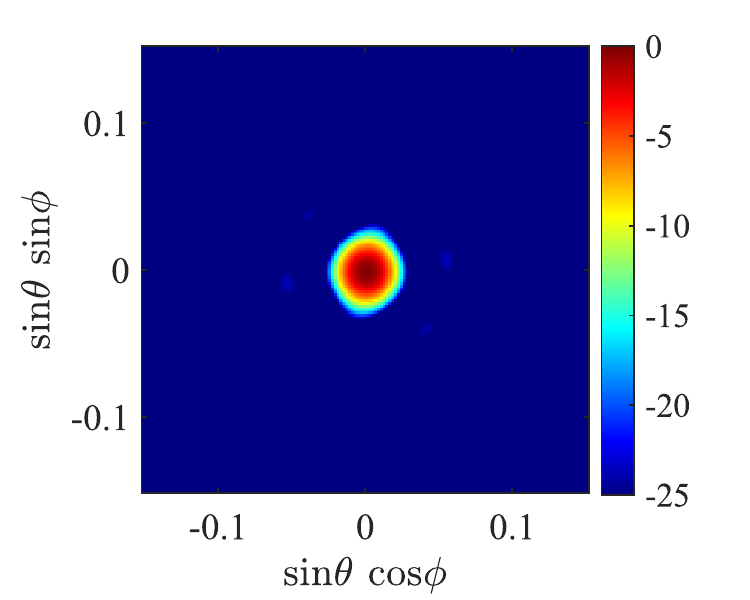}
	\caption{Theoretical PSF of each case in figure~\ref{SimAr}. It can be seen the PSF changes depending on the array formation. The PSF can be used to characterize the imaging system in term of unambiguous FOV, resolution and quality of image reconstruction of the imager.}
	\label{ThePSF}
\end{figure*}

	\begin{figure}[t]
		\noindent
		\centering
		\includegraphics[width=0.4\textwidth]{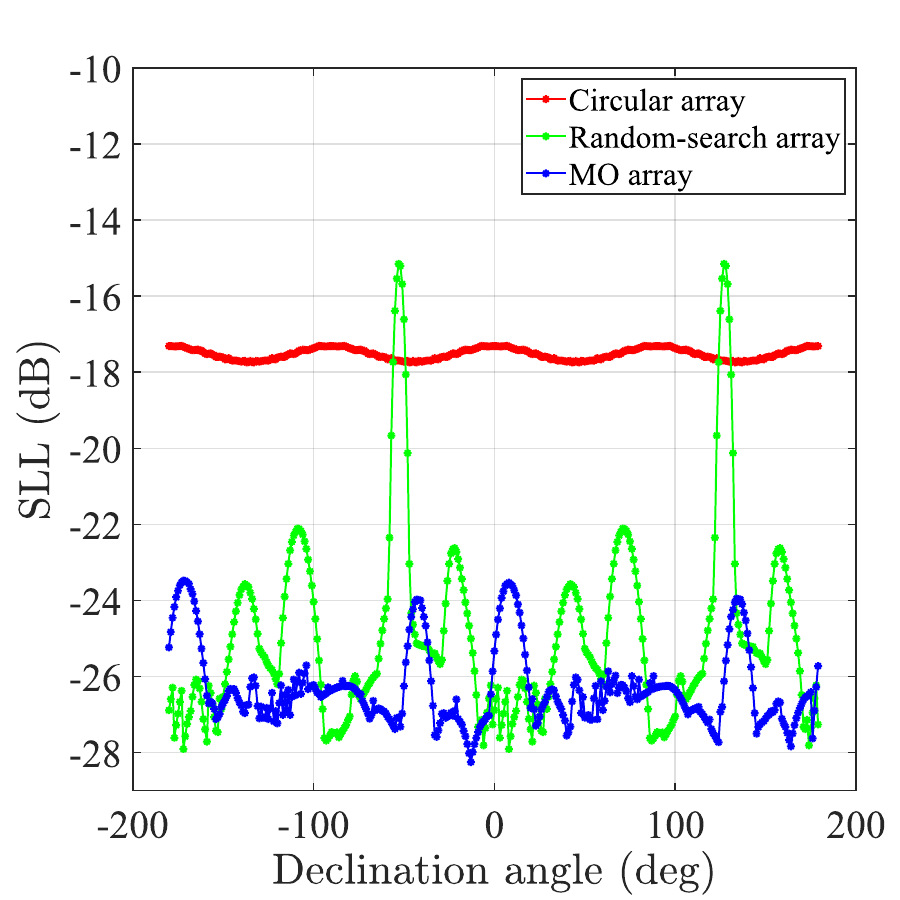}
		
		\caption{Sidelobe level (SLL) versus declination plot. Using as reference the center pixel and strongest point of the PSF we take 360$\degree$ of angle around. This plot shows the comparison of the three array PSF SLL at every angle in increments of 1$\degree$. The behavior shown is expected and it can be appreciated as the SLL in the MO and random-search case is lower, it lowers it by a value of 7~dB(random-search case) 9~dB (MO case).}
		\label{TheoSLL}
	\end{figure}

For this problem, two optimization techniques were employed. 
The first approach was a random-search method, which consisted of randomly selecting six million candidate array configurations and evaluating the number of redundant samples for each case. The resulting array configuration is shown in Fig.~\ref{SimAr}.
The second technique involved the use of a genetic algorithm (GA) to perform multi-objective optimization (MO). To formulate the optimization problem, three objective functions were defined as
\begin{tabbing}
	\hspace*{0.5in} \= Maximize unique $\mathrm{Samples}(n)$\\
	\> Minimize $\mathrm{Resolution}(n)= \theta^{(\alpha,\beta)}_{\mathrm{HPBW}} \approx 0.88 \frac{\lambda}{D_{x,y}}$\\
	\> Maximize $\mathrm{FOV}(n)= \frac{\lambda}{2 d_{x,y}}$\\
	\> Subject to $\mathrm{Sum}(n) \leq 24$,\\
	\> $D_{x,y} \leq 202\,\mathrm{mm}$,\\
	\> $d_{x,y} \geq 26\,\mathrm{mm}$,\\
	\> with $1 \leq n \leq 48$.
\end{tabbing}
The optimization algorithm operates on a vector of 24 elements representing indices corresponding to antenna placement locations shown in Fig.~\ref{pos}. Each antenna is assigned a unique index, and the algorithm uses this index vector to extract the corresponding spatial coordinates. These coordinates are then used to evaluate the objective functions. To determine the number of unique samples, the sampling function is computed and visualized, and each distinct sample is counted once, thereby allowing identification of configurations that maximize sampling diversity. 

The optimization was implemented using the \textit{gamultiobj} solver in MATLAB with the following parameters: population size of 500, 200 generations, a crossover fraction of 0.8, a Pareto fraction of 0.6, and a uniform mutation function. Multiple optimization runs were performed with different population settings, each producing five Pareto-optimal solutions. The final outcome of this optimization process resulted in the third array configuration, shown in Fig.~\ref{SimAr}. This configuration achieved good resolution in both the horizontal and vertical dimensions while maintaining a wide field of view and a large number of unique samples.

\section{Analysis}

To analyze and compare the performance of the proposed array configurations, simulations were conducted for each array geometry. The evaluation focused on image reconstruction quality, spatial resolution, and main-lobe to sidelobe level characteristics. First, the sampling function corresponding to each array formation was computed and is shown in Fig.~\ref{samp}. The number of unique samples generated by each array was determined through cross-correlation of the element positions. The resulting counts of unique samples were 289 for the circular array, 477 for the random-search array, and 545 for the MO-optimized array. Following \eqref{psf}, an inverse Fourier transform was applied to each sampling function to obtain the corresponding theoretical PSF. The normalized PSFs for all array configurations are presented in Fig.~\ref{ThePSF}. As expected, the PSF shape varies across configurations because it is directly determined by the sampling function, which itself depends on the array geometry. This relationship highlights the connection between the number of unique samples and the resulting sidelobe behavior. The sidelobe level (SLL) was computed as a function of angular position relative to the strongest pixel in each normalized PSF. Figure~\ref{TheoSLL} presents the SLL as a function of declination angle, illustrating the PSL for each configuration. 

Table~\ref{TableSLL} summarizes the number of unique samples for each array alongside the corresponding average SLL values in dB. The circular array exhibits an average SLL of $-17.52$~dB, the random-search array achieves $-24.96$~dB, and the multi-objective array attains $-26.29$~dB, representing an improvement of nearly 10~dB between the circular and MO configurations. These results confirm that more unique samples leads to reduced SLL. In the random-search configuration, improved sidelobe performance relative to the circular array is observed; however, the reconstructed PSF exhibits noticeable broadening. This effect arises because the random-search method exclusively optimizes for sampling diversity without explicitly considering resolution or field-of-view constraints. In contrast, the MO-optimized array retains the low sidelobe characteristics achieved by the random-search approach while also preserving strong resolution performance in both the horizontal and vertical dimensions. This outcome is a direct consequence of the multi-objective formulation, which simultaneously optimizes resolution and FOV while minimizing sampling redundancy.

\begin{table}[t!]\caption{Theoretical PSF Unique Samples and Average SLL}\label{TableSLL}
	\resizebox{\columnwidth}{!}{%
		\begin{tabular}{c|c|c}
			\hline
			& \# Unique Samples & Avg SLL (dB) \\ \hline
			Circular & 289 & -17.52 \\ \hline
			Random-search & 477 & -24.96 \\ \hline
			MO & 545 & -26.29 \\ \hline
		\end{tabular}%
	}
\end{table}

\begin{table}[t!]\caption{SSIM Values of the Example Scenes}\label{ssimtab}
	\resizebox{\columnwidth}{!}{
		
		\begin{tabular}{c|c|c|c|c}
			\hline
			& S1 & S2 & S3 & S4 \\ \hline
			Circular & 0.125 & 0.077 & 0.079 & 0.097 \\ \hline
			Random-search & 0.135 & 0.111 & 0.099 & 0.100 \\ \hline
			MO & 0.142 & 0.138 & 0.107 & 0.112 \\ \hline
		\end{tabular}
		
	}
\end{table}

	
	

\begin{figure*}[t!]
	\noindent
	\includegraphics[width=1\textwidth]{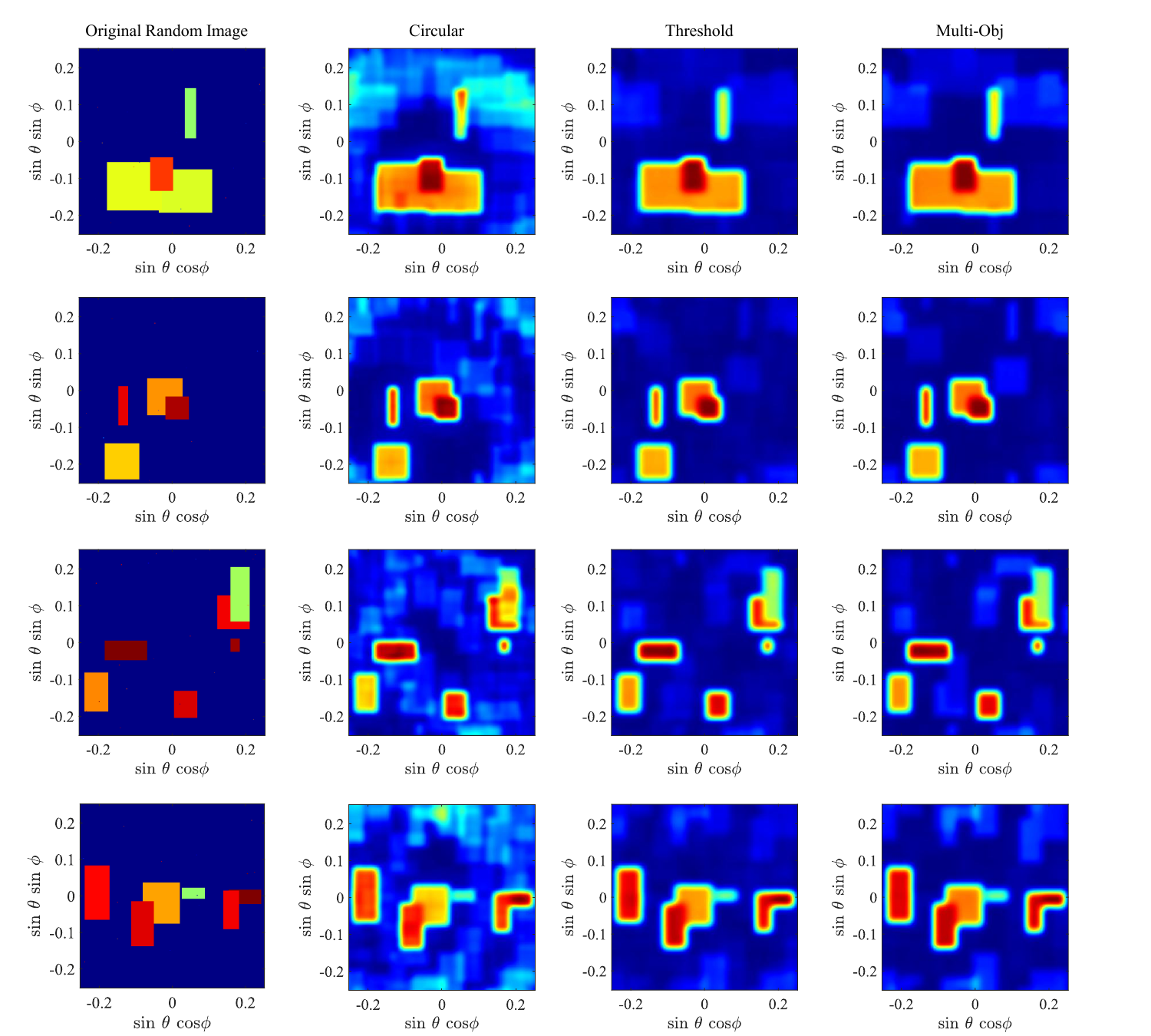}
	
	\hfil
	\caption{From left to right, the reconstructed images corresponding to each array configuration are presented. From top to bottom, scenes S1 through S4 are shown, representing the different randomly generated test cases selected for comparison and SSIM analysis. The MO-optimized array demonstrates consistent performance across all scenarios, producing cleaner reconstructions with reduced sidelobe artifacts. In contrast, the circular array exhibits more pronounced sidelobes, which introduce visible clutter and degrade image quality. The random-search array maintains reliable reconstruction performance; however, its lower resolution leads to slight image blurring, and in some cases, aliasing effects begin to appear within the reconstructed targets.}
	\label{SSIMim}
\end{figure*}

A simulated set of random millimeter-wave test scenes were generated to facilitate an analysis of the image reconstruction performance of the arrays, as shown in Fig.~\ref{SSIMim}. The theoretical PSF associated with each configuration was then used to simulate the reconstructed scene via convolution with the scene. The reconstructed images were subsequently compared with the original scenes using the SSIM, which quantifies structural agreement between two images, with values ranging from $-1$ (completely decorrelated) to $1$ (perfect reconstruction)~\cite{1284395}. The SSIM results obtained from the random scenes provide insight into how effectively each array captures spatial information. Scenes dominated by low spatial frequency content generally yield higher SSIM values, as the arrays more accurately capture broad intensity variations. In contrast, scenes containing high spatial frequency features, such as small or closely spaced targets, place greater demands on array resolution and more clearly reveal sidelobe effects in the reconstructed images.

Four random test scenes were selected for evaluation. The randomness was introduced in both the geometric shapes and their corresponding intensity levels in order to assess how effectively each array captures spatial information and to enable a fair performance comparison. 
The scenes were created to extend beyond of the unambiguous FOV to analyze the effect of aliasing in the images. Fig.~\ref{SSIMim} presents the reconstructed scenes, labeled S1, S2, S3, and S4. After reconstruction, the cropped unambiguous FOV was compared to the original scene to compute the SSIM values reported in Table~\ref{ssimtab}. In all cases, the MO-optimized array achieved the highest SSIM values. The random-search array consistently improved upon the circular configuration, but did not match the performance of the MO design. The circular array benefits from wide baselines distributed over many angles, which can provide strong angular resolution. However, the reconstructed images exhibited noticeable sidelobe contamination, resulting in visually noisy or cluttered scenes.

In scenes S1 and S2, aliasing artifacts were strongly evident within the unambiguous FOV. In practical measurements, such artifacts could obscure weaker targets or make detection more challenging. Both the random-search and MO-optimized configurations produced visibly cleaner reconstructions with reduced ambiguities. The primary distinction between these two designs was that the MO-optimized array achieved higher resolution while also minimizing sample redundancy. As a result, it produces sharper and cleaner images compared to the other configurations. These results demonstrate that the MO-optimized array provided a better balance between resolution, sidelobe suppression, and ambiguity reduction, making it better suited for reliable target detection and separation. Analysis of the reconstruction results indicates that array configurations with a larger number of unique spatial samples produce cleaner main lobes and reduced sidelobe artifacts. This behavior is reflected in consistently higher SSIM values across the evaluated scenes. In particular, the MO-optimized array demonstrates the strongest overall performance, providing improved structural fidelity while maintaining the capability to resolve fine scene details within the unambiguous FOV. These findings suggest that the MO array design enhances practical imaging performance by effectively balancing the trade-off between spatial resolution and sidelobe suppression across a diverse set of randomly generated test scenes.

\begin{figure}[t!]
	\noindent
	\centering
	\includegraphics[width=0.55\textwidth]{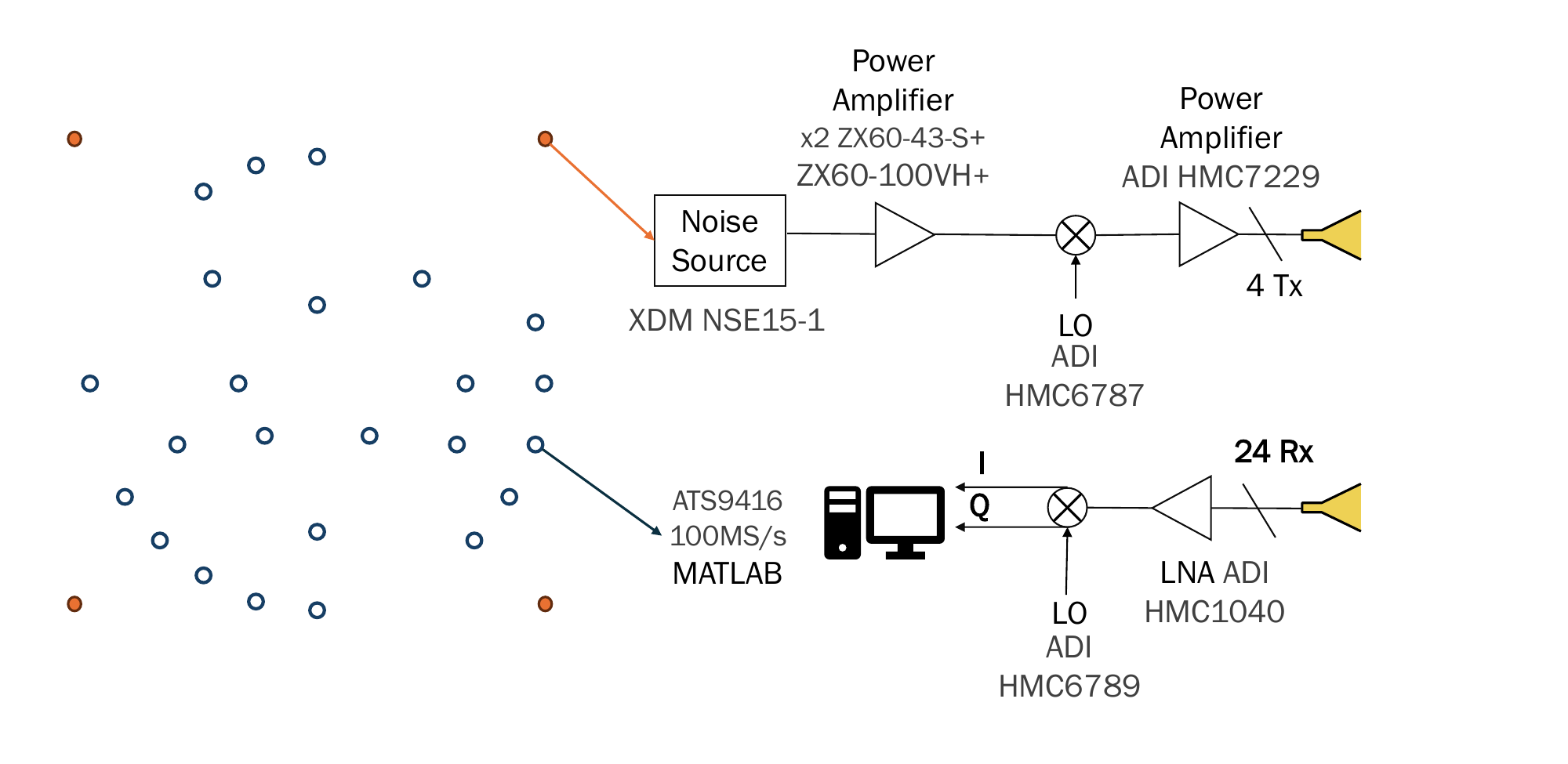}
	
	\caption{Block diagram of the transmitter and receiver chains.}
	\label{Schem}
\end{figure}

\begin{figure*}[t!] 
	\centering
	\includegraphics[width=0.3\textwidth]{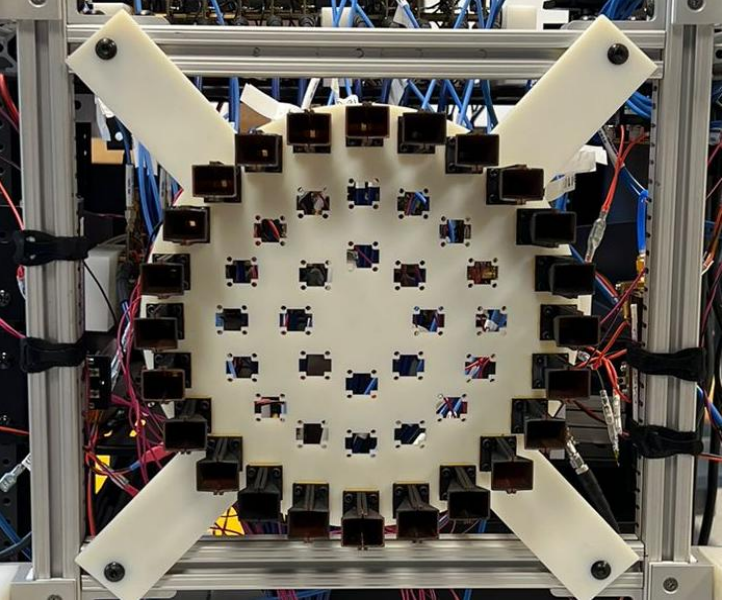}
	\hfill
	\includegraphics[width=0.3\textwidth]{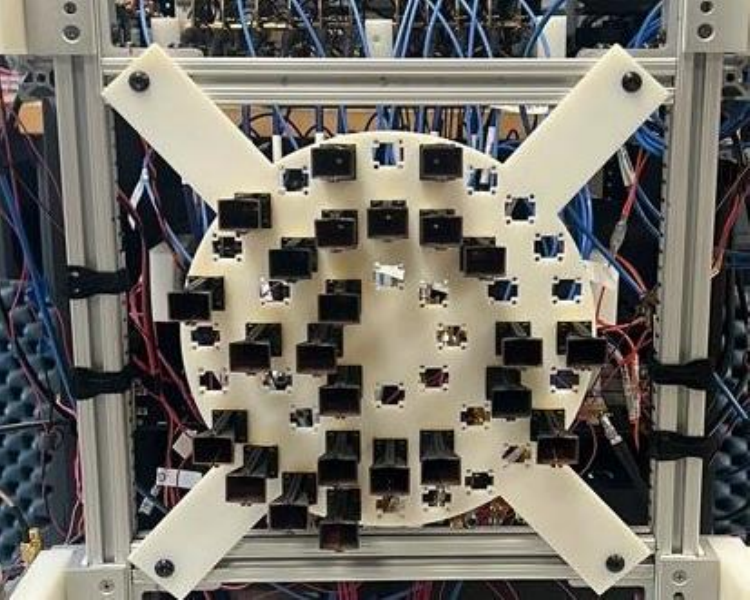}
	\hfill
	\includegraphics[width=0.3\textwidth]{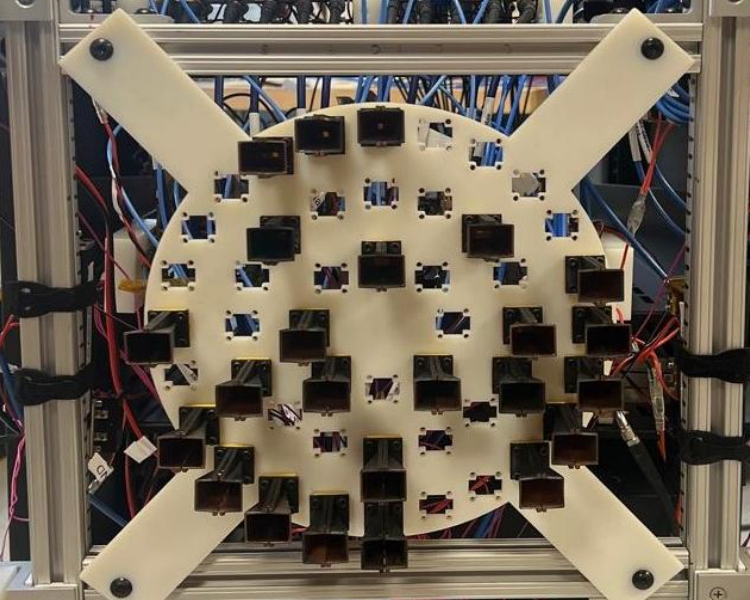}
	\caption{Experimental array layouts for the circular array (left), random-search array (middle) and multi-objective optimized array (right).}
	\label{PArr}
\end{figure*}

\begin{figure*}[t] 
	\centering
	\includegraphics[width=0.3\textwidth]{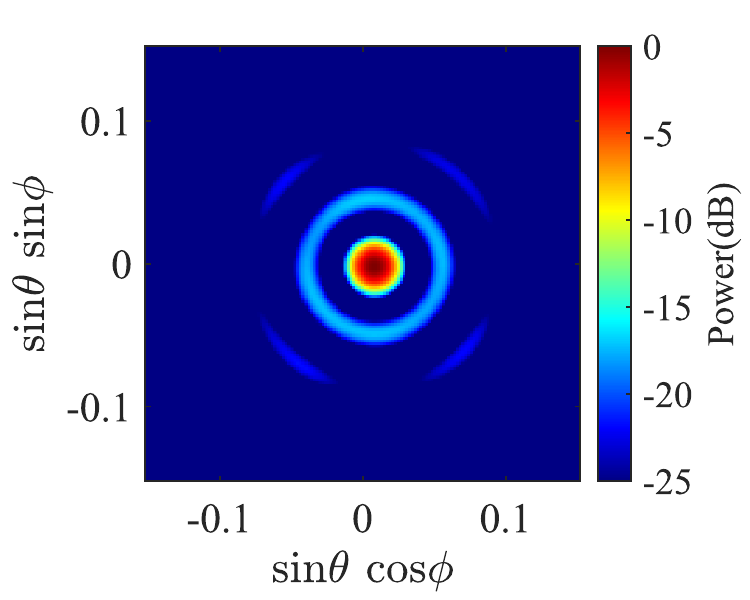}
	\hfill
	\includegraphics[width=0.3\textwidth]{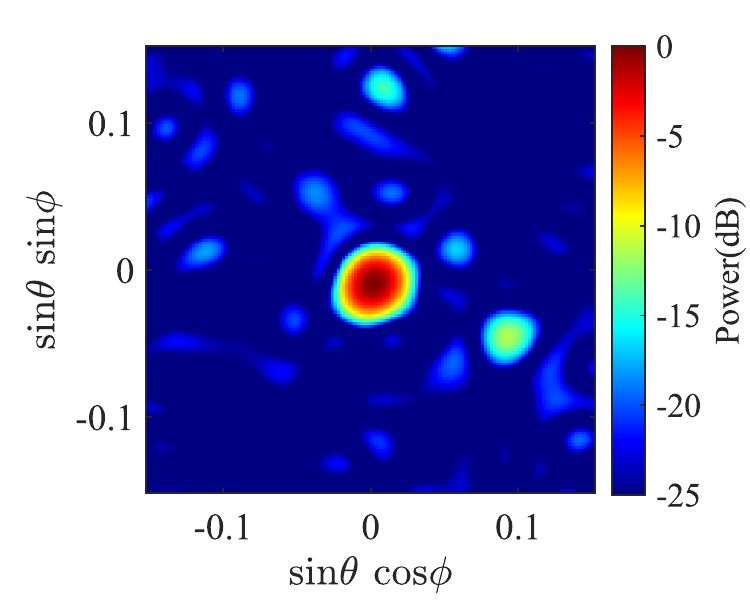}
	\hfill
	\includegraphics[width=0.3\textwidth]{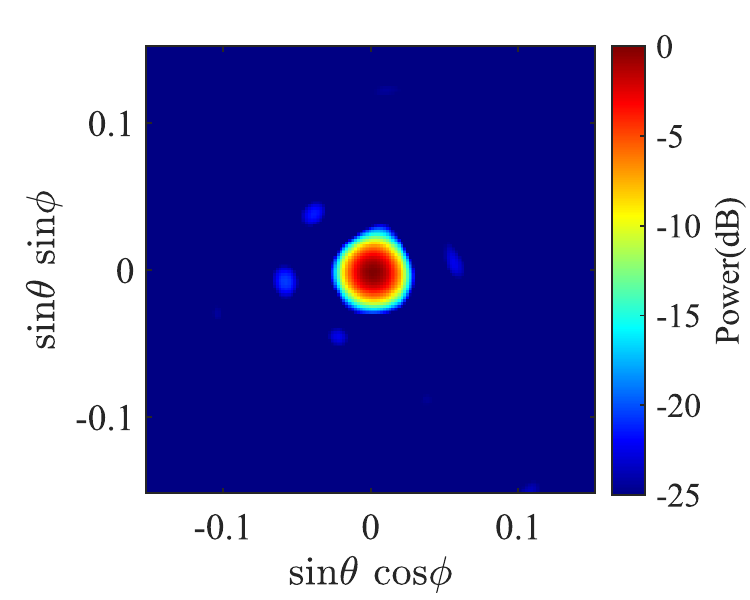}
	\caption{Measured point spread functions for the circular array (left), random-search array (middle) and multi-objective optimized array (right).}
	\label{cal}
\end{figure*}

\begin{figure}[!h]
	\noindent
	\centering
	\includegraphics[width=0.4\textwidth]{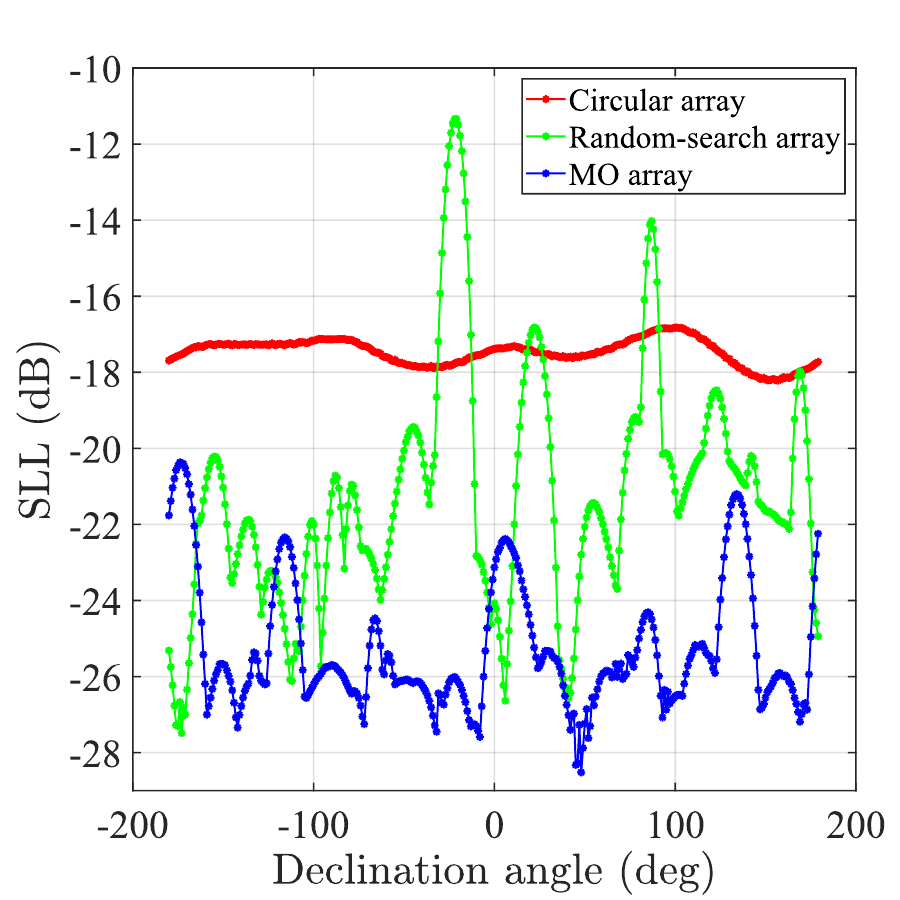}
	
	\caption{Measured Sidelobe level (SLL) versus declination plot. The behavior shown is expected from what was obtained in theory, showing the calibration is close to theory and also showing avg SLL of -17.46 dB (circular), -21.25 dB (random-search) and -25.28 dB (MO).}
\end{figure}


\section{Experimental System Design and Calibration}

In order to demonstrate the performance of the array formation, a set of experiments were performed using a 38-GHz system consisting of four transmitters illuminating the scene with noise waveforms mimicking thermally-generated random signals. A block diagram can be seen in Fig.~\ref{Schem}. The system consisted of a 24-channel receiving array along with four noise transmitters that were placed at a wider baseline than the widest baseline of the receiving array to ensure that the signals incident on the scene were spatially incoherent. Since the transmitter noise sources were independent, the signals incident on the scene were therefore spatially and temporally incoherent. Each transmitter used a separate baseband noise source, the signal from which was upconverted to the millimeter-wave carrier frequency using Analog Devices (ADI) HMC6787 upconverters. Before transmission, the millimeter-wave noise signals were amplified using ADI HMC7229 power amplifiers. The carrier frequency was generated by a 19-GHz local oscillator (LO) using a Keysight N5183A. The upconverters included an integrated frequency doubler that converts the 19-GHz LO to the desired 38-GHz carrier frequency. The signals captured at the receiving array were amplified using ADI HMC1040 low-noise amplifiers and then quadrature downconverted using ADI HMC6789 downconverters. All 48 I/Q baseband signals were digitized by three ATS9416 samplers that were hosted in a computer. The sample rate on each channel was 100~MS/s. All signal processing and image reconstruction was implemented in a host computer using MATLAB.
	
The array frame was fabricated and integrated into the imaging system, and the antenna placement was reconfigured to match each array geometry under evaluation. Optical photographs of the implemented array configurations are shown in Fig.~\ref{PArr}. Prior to imaging experiments, system calibration is required to ensure that relative static phase offsets in the receiving channels are corrected. Previous work introduced a calibration approach based on redundant baselines within the antenna array, which naturally produce redundant sampling points~\cite{9079644}. This requirement influenced array formation, but favored geometries with baseline redundancy. Because the arrays in this work are focused on minimizing redundancy, a different calibration approach was required. By employing an active antenna transmitting a continuous-wave signal at 38~GHz as a test target, a direct calibration procedure was implemented following~\cite{90000}.
Because the calibration target is a known point source, linearization of both amplitude and phase responses can be applied, resulting in a set of complex calibration weights for each receive element. Specifically, the spatial locations of antenna pairs are used to derive weights that compensate amplitude and phase mismatches across the receiver channels. These weights enable accurate reconstruction of the point target generated by the single-tone active antenna positioned at the center of the array field of view. This calibration approach enables investigation of array geometries with reduced baseline redundancy while preserving accurate imaging performance. The calibrated point-source responses for each array configuration are presented in Fig.~\ref{cal}. All three cases exhibit structures consistent with their corresponding theoretical PSFs shown in Fig.~\ref{ThePSF}, validating the calibration process and confirming the expected dependence of the measured response on array geometry.

\section{Experiments}

\begin{table}[t!]\caption{Measured PSF Unique Samples and Average SLL}
	\resizebox{\columnwidth}{!}{%
		\begin{tabular}{c|c|c}
			\hline
			& \# Unique Samples & Avg SLL (dB) \\ \hline
			Circular & 289 & -17.46 \\ \hline
			Random-search & 477 & -21.25 \\ \hline
			MO & 545 & -25.28 \\ \hline
		\end{tabular}%
	}
\end{table}

\begin{figure*}[t!]
	\noindent
	\includegraphics[width=1\textwidth]{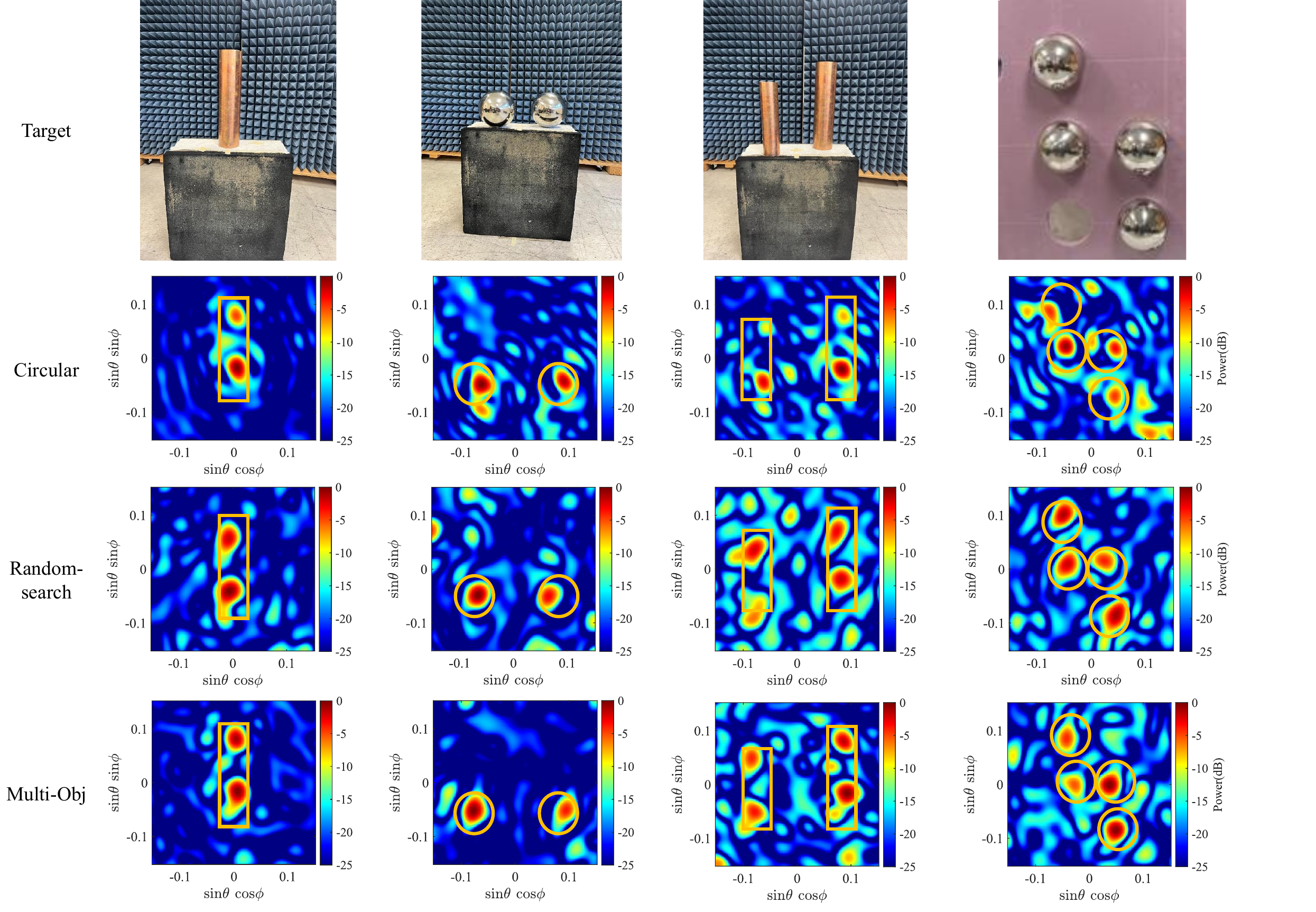}

	\caption{Four targets were reconstructed and the performance was compared. The first target is a copper tape cylinder that measures 0.47~m of height and a diameter of 0.102~m, it was placed at the center of the FOV, 1.88~m away from the system. Second target consists of two spheres with diameter equal to 0.203~m and RCS of -14.89~dbsm, the left sphere was moved 0.127~m to the left of the FOV center and at 1.83~m distance from the system. The sphere on the right was moved 0.165~m to the right and is 1.96~m away from the system. The third target utilizes the same distances to the right and left and back and forward from the second target, in this case the cylinder used on the right is the same from the first target, the cylinder to the left had a height of 0.37~m and a diameter of 0.076~m. Finally the last target consist of four spheres with diameter of 10~cm and distance between spheres of 0.15~m with -21~dbsm RCS. The target was placed at 1.524~m from the system.}
	\label{results}
\end{figure*}

Four experimental targets were reconstructed to evaluate and compare array performance. The first target was a copper-tape cylinder with a height of 0.47~m and a diameter of 0.10~m, positioned at the center of the field of view (FOV) at a distance of 1.88~m from the system. The second target consisted of two spheres, each with a diameter of 0.20~m and a radar cross section (RCS) of $-15$~dBsm. The left sphere was positioned 0.13~m to the left of the FOV center at a distance of 1.83~m, while the right sphere was located 0.17~m to the right and 1.96~m from the system. The third target maintained the same lateral and depth offsets as the second configuration but replaced the spheres with cylindrical objects. The right cylinder was identical to the cylinder used in the first target, whereas the left cylinder had a height of 0.37~m and a diameter of 0.08~m. The fourth target consisted of four spheres with diameters of 0.1~m, separated by 0.15~m center-to-center, each exhibiting an RCS of approximately $-21$~dBsm. This target arrangement was positioned 1.52~m from the system.

The reconstructed images for all cases are presented in Fig.~\ref{results}. The three array configurations successfully reconstructed the first three targets, while noticeable differences emerged in the fourth scenario. For the first target, the circular array produced a reconstruction with relatively fine resolution; however, some target information was diminished compared to the random-search and MO configurations. Consistent with simulation results, sidelobe levels decreased as the number of redundant samples was reduced, leading to improved image clarity. 

In the second scenario, both spheres were reconstructed across all arrays. Notably, the random-search and MO configurations more accurately captured the amplitude difference between the two spheres, particularly for the right sphere located farther from the array, indicating improved scene representation. The third scenario, involving two cylinders, was also successfully reconstructed in all cases, with observable differences corresponding to variations in cylinder height and range. The fourth target highlighted the most pronounced performance differences. The circular array struggled to fully resolve the four-sphere configuration due to sidelobe interference that obscured portions of the scene. The random-search array achieved clearer reconstruction of the targets, while the MO-optimized array provided the most reliable result, enabling straightforward identification of the central targets within the FOV. This improved performance is attributed to reduced sidelobe levels and stronger main-lobe dominance, allowing the targets to appear more distinctly in the reconstructed image.

	
	
	


	\section{Conclusion}
This work investigated the impact of receive array geometry on reconstruction performance in a 24-element 38-GHz AIM imaging system. Three array configurations were analyzed: a circular layout, a random-search optimized array designed to maximize unique spatial frequency samples, and a MO-optimized array that simultaneously considered spatial sampling diversity, system resolution, and field of view constraints. Theoretical analysis demonstrated that increasing the number of unique spatial frequency samples improves the point spread function characteristics by reducing sidelobe levels and enhancing spatial frequency coverage. Among the evaluated configurations, the MO-optimized array achieved the highest number of unique samples and produced the lowest theoretical peak sidelobe levels while maintaining strong resolution performance. Simulated image reconstruction using randomly generated scenes confirmed that arrays with greater sampling diversity yield improved structural fidelity, with the MO configuration consistently achieving the highest SSIM values. Experimental validation was performed using a 38-GHz AIM imaging platform with a reconfigurable 24-element receive array implemented within a 48-position aperture. Measured PSFs closely matched theoretical predictions, confirming the effectiveness of the calibration procedure and the expected influence of array geometry on imaging behavior. Imaging experiments with several physical targets demonstrated that the optimized arrays produced clearer reconstructions and reduced sidelobe artifacts compared to the circular configuration.  Array layout optimization can significantly improve reconstruction quality without increasing hardware complexity. While this work is applied to a restricted antenna array in an imaging system with spatial constraints, the approach can be directly extended to other Fourier domain array design problems.                          	
	
\addtolength{\textheight}{-3cm}

\bibliographystyle{IEEEtran}
\bibliography{IEEEabrv,reference_bib_2}

\begin{thebibliography}{10}
\providecommand{\url}[1]{#1}
\csname url@samestyle\endcsname
\providecommand{\newblock}{\relax}
\providecommand{\bibinfo}[2]{#2}
\providecommand{\BIBentrySTDinterwordspacing}{\spaceskip=0pt\relax}
\providecommand{\BIBentryALTinterwordstretchfactor}{4}
\providecommand{\BIBentryALTinterwordspacing}{\spaceskip=\fontdimen2\font plus
\BIBentryALTinterwordstretchfactor\fontdimen3\font minus
  \fontdimen4\font\relax}
\providecommand{\BIBforeignlanguage}[2]{{%
\expandafter\ifx\csname l@#1\endcsname\relax
\typeout{** WARNING: IEEEtran.bst: No hyphenation pattern has been}%
\typeout{** loaded for the language `#1'. Using the pattern for}%
\typeout{** the default language instead.}%
\else
\language=\csname l@#1\endcsname
\fi
#2}}
\providecommand{\BIBdecl}{\relax}
\BIBdecl

\bibitem{942570}
D.~M. Sheen, D.~L. McMakin, and T.~E. Hall, ``Three-dimensional millimeter-wave
  imaging for concealed weapon detection,'' \emph{IEEE Trans. Microw. Theory
  Techn.}, vol.~49, no.~9, pp. 1581--1592, Sep. 2001.

\bibitem{yujiri2003passive}
L.~Yujiri, M.~Shoucri, and P.~Moffa, ``Passive millimeter wave imaging,''
  \emph{IEEE Microw. Mag.}, vol.~4, no.~3, pp. 39--50, 2003.

\bibitem{4337827}
R.~Appleby and R.~N. Anderton, ``Millimeter-wave and submillimeter-wave imaging
  for security and surveillance,'' \emph{Proc. IEEE}, vol.~95, no.~8, pp.
  1683--1690, Aug 2007.

\bibitem{11236959}
D.~Chen and J.~A. Nanzer, ``Space–time dynamic antenna array design for
  fourier-domain microwave remote sensing,'' \emph{IEEE Transactions on
  Geoscience and Remote Sensing}, vol.~63, pp. 1--17, 2025.

\bibitem{ulander1998ultra}
L.~M. Ulander and P.-O. Fr{\"o}lind, ``Ultra-wideband sar interferometry,''
  \emph{IEEE Transactions on Geoscience and Remote Sensing}, vol.~36, no.~5,
  pp. 1540--1550, 1998.

\bibitem{rau2011multi}
U.~Rau and T.~J. Cornwell, ``A multi-scale multi-frequency deconvolution
  algorithm for synthesis imaging in radio interferometry,'' \emph{Astronomy \&
  Astrophysics}, vol. 532, p. A71, 2011.

\bibitem{diebold2021unified}
A.~V. Diebold, T.~Fromenteze, E.~Kpr{\'e}, C.~Decroze, M.~F. Imani, and D.~R.
  Smith, ``Unified reciprocal space processing for short-range active and
  passive imaging systems,'' \emph{IEEE Open Journal of Antennas and
  Propagation}, vol.~3, pp. 124--134, 2021.

\bibitem{11271319}
J.~R. Colon-Berrios and J.~A. Nanzer, ``Frequency-diverse additive processing
  for active incoherent millimeter-wave imaging,'' \emph{IEEE Open Journal of
  Antennas and Propagation}, vol.~7, no.~1, pp. 184--193, 2026.

\bibitem{8458190}
S.~Vakalis and J.~A. Nanzer, ``Microwave imaging using noise signals,''
  \emph{IEEE Transactions on Microwave Theory and Techniques}, vol.~66, no.~12,
  pp. 5842--5851, 2018.

\bibitem{8904834}
S.~Vakalis, L.~Gong, J.~Papapolymerou, and J.~A. Nanzer, ``40-ghz active
  interferometric imaging with noise transmitters,'' in \emph{2019 16th
  European Radar Conference (EuRAD)}, 2019, pp. 273--276.

\bibitem{Thompson2001}
A.~R. Thompson, J.~M. Moran, and G.~W. Swenson, \emph{Interferometry and
  Synthesis in Radio Astronomy}.\hskip 1em plus 0.5em minus 0.4em\relax John
  Wiley and Sons, 2001.

\bibitem{8861700}
S.~Ellison, S.~Vakalis, and J.~A. Nanzer, ``Optimizing for spatial frequency
  coverage vs. point-spread function sidelobe level in active incoherent
  microwave imaging arrays,'' in \emph{2019 USNC-URSI Radio Science Meeting
  (Joint with AP-S Symposium)}, 2019, pp. 103--104.

\bibitem{9624943}
A.~M. Molaei, S.~Hu, V.~Skouroliakou, V.~Fusco, X.~Chen, and O.~Yurduseven,
  ``Fourier compatible near-field multiple-input multiple-output terahertz
  imaging with sparse non-uniform apertures,'' \emph{IEEE Access}, vol.~9, pp.
  157\,278--157\,294, 2021.

\bibitem{4512143}
N.~Jin and Y.~Rahmat-Samii, ``Analysis and particle swarm optimization of
  correlator antenna arrays for radio astronomy applications,'' \emph{IEEE
  Transactions on Antennas and Propagation}, vol.~56, no.~5, pp. 1269--1279,
  2008.

\bibitem{5525009}
K.~Chao, Z.~Zhao, Z.~Wu, and R.~Lang, ``Application of the differential
  evolution algorithm to the optimization of two-dimensional synthetic aperture
  microwave radiometer circle array,'' in \emph{2010 International Conference
  on Microwave and Millimeter Wave Technology}, 2010, pp. 1212--1215.

\bibitem{9769540}
V.~Skouroliakou, A.~M. Molaei, V.~Fusco, and O.~Yurduseven, ``Fourier-based
  radar processing for multistatic millimetre-wave imaging with sparse
  apertures,'' in \emph{2022 16th European Conference on Antennas and
  Propagation (EuCAP)}, 2022, pp. 1--5.

\bibitem{768786}
F.~Ares-Pena, J.~Rodriguez-Gonzalez, E.~Villanueva-Lopez, and S.~Rengarajan,
  ``Genetic algorithms in the design and optimization of antenna array
  patterns,'' \emph{IEEE Transactions on Antennas and Propagation}, vol.~47,
  no.~3, pp. 506--510, 1999.

\bibitem{0749d8e86a124a648470521f844f7ec1}
M.~Bray, D.~Werner, D.~Boeringer, and D.~Machuga, ``\BIBforeignlanguage{English
  (US)}{Optimization of thinned aperiodic linear phased arrays using genetic
  algorithms to reduce grating lobes during scanning},''
  \emph{\BIBforeignlanguage{English (US)}{IEEE Transactions on Antennas and
  Propagation}}, vol.~50, no.~12, pp. 1732--1742, Dec. 2002.

\bibitem{7305316}
H.~Tsutsumi, Y.~Kuwahara, and H.~Kamo, ``Design of the series fed microstrip
  patch planar array antenna by the parato genetic algorithm,'' in \emph{2015
  IEEE International Symposium on Antennas and Propagation \& USNC/URSI
  National Radio Science Meeting}, 2015, pp. 1854--1855.

\bibitem{10416353}
M.~W. Wolff and J.~A. Nanzer, ``Application of pseudoweights in antenna array
  optimization and design,'' \emph{IEEE Antennas and Wireless Propagation
  Letters}, vol.~23, no.~5, pp. 1478--1482, 2024.

\bibitem{996017}
K.~Deb, A.~Pratap, S.~Agarwal, and T.~Meyarivan, ``A fast and elitist
  multiobjective genetic algorithm: Nsga-ii,'' \emph{IEEE Transactions on
  Evolutionary Computation}, vol.~6, no.~2, pp. 182--197, 2002.

\bibitem{9127123}
S.~Vakalis, D.~Chen, and J.~A. Nanzer, ``Toward space–time incoherent
  transmitter design for millimeter-wave imaging,'' \emph{IEEE Antennas and
  Wireless Propagation Letters}, vol.~19, no.~9, pp. 1471--1475, 2020.

\bibitem{4738328}
M.~Karaman, I.~O. Wygant, O.~Oralkan, and B.~T. Khuri-Yakub, ``Minimally
  redundant 2-d array designs for 3-d medical ultrasound imaging,'' \emph{IEEE
  Transactions on Medical Imaging}, vol.~28, no.~7, pp. 1051--1061, 2009.

\bibitem{9079644}
S.~Vakalis, L.~Gong, Y.~He, J.~Papapolymerou, and J.~A. Nanzer, ``Experimental
  demonstration and calibration of a 16-element active incoherent
  millimeter-wave imaging array,'' \emph{IEEE Transactions on Microwave Theory
  and Techniques}, vol.~68, no.~9, pp. 3804--3813, 2020.

\bibitem{9753682}
J.~R. Colon-Berrios, S.~Vakalis, D.~Chen, and J.~A. Nanzer, ``Incoherent point
  spread function estimation and multipoint deconvolution for active incoherent
  millimeter-wave imaging,'' \emph{IEEE Microwave and Wireless Components
  Letters}, vol.~32, no.~6, pp. 800--803, 2022.

\bibitem{Nanzer2012}
J.~A. Nanzer, \emph{Microwave and Millimeter-Wave Remote Sensing for Security
  Applications}.\hskip 1em plus 0.5em minus 0.4em\relax Artech House, 2012.

\bibitem{1284395}
Z.~Wang, A.~Bovik, H.~Sheikh, and E.~Simoncelli, ``Image quality assessment:
  from error visibility to structural similarity,'' \emph{IEEE Transactions on
  Image Processing}, vol.~13, no.~4, pp. 600--612, 2004.

\bibitem{90000}
C.~L.~C. G.B~Taylor and R.~Perley, ``Synthesis imaging in radio astronomy ii,
  chapter 5,'' \emph{ASP Conference Series}, vol. 180, pp. 79--110, 1999.

\end{thebibliography}

\end{document}